\documentclass[twocolumn,prl,aps,showpacs]{revtex4}
\usepackage{epsfig}
\def\be{\begin{equation}}
\def\ee{\end{equation}}

\begin{document}
\title{Quantum Step Heights in 
Hysteresis Loops of Molecular Magnets}
\author{Jie Liu$^{1}$, Biao Wu$^1$, Li-Bin Fu$^2$,
Roberto B. Diener$^{1}$, and Qian Niu$^1$}
\affiliation{$^1$Department of Physics, The 
University of Texas, Austin, Texas 78712}
\affiliation{$^2$Institute of Applied Physics and Computational 
Mathematics,\\P.O.Box  8009, Beijing 100088, China}

\date{Nov. 15th, 2001}

\begin{abstract}
We present an analytical theory on the heights of the quantum steps
observed in the hysteresis loops of molecular magnets.
By considering the dipolar interaction between
molecular spins, our theory successfully yields the step heights
measured in experiments, and  reveals a scaling law 
for the  dependence of the heights on the sweeping rates
hidden in the experiment data on Fe$_8$ and Mn$_4$.
With this theory, we show  how to accurately determine the tunnel
splitting of a single molecular spin from the step heights. 
\end{abstract}

\maketitle

Crystals of molecular magnets, such as Fe$_8$ and Mn$_{12}$,
have attracted much attention for their connection to macroscopic 
quantum tunneling and Berry phases\cite{mn12,ljk,all,lzsci}.  
They may also have important applications in magnetic memory and 
quantum computing \cite{lzsci,qc}. The earliest and most spectacular 
observation on such a system is the quantum steps 
in the hysteresis loop of magnetization
at low temperatures \cite{mn12}. 

These quantum steps are a manifestation of macroscopic quantum tunneling,
resulting from the tunneling between different spin states of large 
molecular spins ($S=10$ for both Fe$_8$ and Mn$_{12}$, $S=9/2$ for Mn$_4$).
As in any many-spin system, this tunneling phenomenon is naturally 
complicated by the interaction between spins and  other environmental effects.
A full understanding of this phenomenon demands a successful theory 
on the quantum steps, describing their widths, shapes, and heights.
So far, besides speculations, no serious attempt has been made
to develop quantitative theories on any of the step features. 

\begin{figure}[!htb]
\begin{center}
\resizebox *{8.0cm}{6.18cm}{\includegraphics*{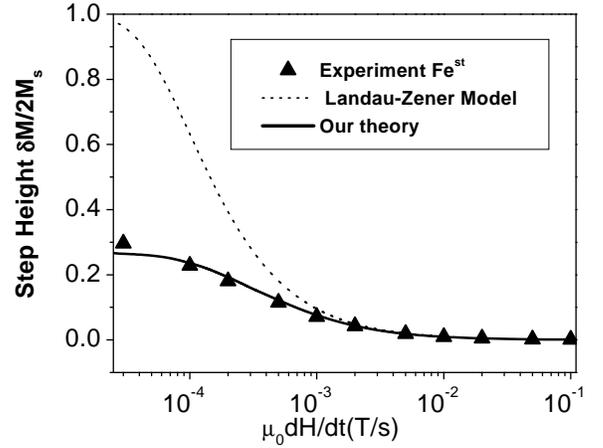}}
\end{center}
\caption{Comparison between the experimental data on 
Fe$_8$\cite{w1}, the LZ model, and our theory. 
The normalized step height $\delta M/2M_s$ 
is the final fraction of up-spins $F_{fin}$ after one sweep.}
\label{fig:split}
\end{figure}

In this letter we present a successful theory on the step height
as the first step towards a complete theory on the quantum
steps in the hysteresis loop. Since the height measures the 
tunneling probability between different spin states, it is the 
most prominent feature of the quantum step, and holds the key to 
understanding the underlying tunneling dynamics.
In Fig.\ref{fig:split},  we have adapted the experimental data on Fe$_8$
from Ref.\cite{w1}, and show how the step height 
between spin states $S_z=\pm 10$ changes with the sweeping rates. 
The data are compared to the Landau-Zener (LZ) model\cite{lz},
which has been used to extract the tunnel splitting $\Delta$
of a single molecular spin from the step heights\cite{lzsci,newexp}.
When we fit the LZ model with the data at the fast sweeping regime 
there is a dramatic difference at slow sweepings:
a two-third suppression.

By taking into account the dipolar interaction between 
molecular spins, our theory successfully gives
the step heights measured in the experiment, as shown 
in Fig\,.\ref{fig:split}. Two physical 
mechanisms influencing step heights are identified:
the spin shuffling in the evolving 
distribution of dipolar fields and the dipolar 
interaction between spins in the resonance window.
Furthermore, our theory reveals an $\alpha/\Delta^2$ scaling
law for the  dependence of the heights on the sweeping rate $\alpha$ of
the external field. This law is confirmed by the collapse
of the experimental data in terms of the scaled sweeping rates 
(see Fig. \ref{fig:scaling}). As a direct application of our theory, we show
how to accurately extract the tunnel splitting $\Delta$ of a 
single molecular spin from the step heights
and the sample geometry. In our theory there are no
adjustable parameters.

We argue that nuclear spins do not appreciably 
affect the tunneling dynamics under a fast sweeping field
except for a renormalization of the tunnel splitting.
In the relaxation experiments \cite{exp} where the 
external field remains constant, the tunneling is strongly affected 
by the nuclear spins as recognized by Prokof'ev and Stamp \cite{stamp}. 
However, in the sweeping field experiments, the role of nuclear spins
is marginalized by the sweeping fields. This can be clearly seen in 
Fig. 6 of Ref. \cite{w2}, where the relaxation with a constant external 
field is shown to be much slower than  the one with a sweeping field 
of the slowest rate applied $0.04$mT/s.

We consider spin lattices, such as crystals of
Fe$_8$, Mn$_{12}$, and Mn$_4$, in which the spins interact with each other
through the dipolar potential, $d(\vec{r})=E_D(1-3\cos^2\theta)\Omega_0/r^3$, 
where $\vec{r}$ is the displacement vector between the spins, $\theta$ is
the angle between $\vec{r}$ and the easy axis,
$\Omega_0$ is the unit cell volume, and 
$E_D=\frac{2\mu_0}{4\pi}(Sg\mu_B)^2/\Omega_0$ gives the interaction
strength. Our theory will be compared to the experiment, mainly 
on crystals of Fe$_8$ where the experimental data on step heights are 
the most abundant \cite{w1,w2}. For simplicity, we focus on one step, 
that is, the tunneling between two spin states (for example, 
$S_z=\pm 10$ for Fe$_8$); it is rather straightforward
to extend our theory to study multi-step tunneling.

We now have a system of Ising spins sitting at each site of
a lattice. In a sweeping magnetic field along the easy axis, 
the spins will flip from one state to the other back and forth 
as a result of the tunneling driven by the sweeping.
However, at any given moment, only a small fraction of 
spins are flipping by being in the resonance window while the others 
remain static. This can be understood by first considering an isolated
spin in a sweeping field, which can be described exactly
with the LZ model. In the LZ model the flipping occurs mainly in a 
tunneling time interval,
when the Zeeman energy bias $\gamma=2g\mu_BS\mu_0H$\cite{ss} between the 
two spin states caused by the changing external field 
becomes very small, $|\gamma|\le \Delta_{win}/2$. This tunneling
time defines the resonance window, whose width $\Delta_{win}$ is 
the tunnel splitting $\Delta$ at the adiabatic limit and 
$\sqrt{2\alpha}$ ($\alpha=\hbar{{\rm d}\gamma\over {\rm d}t}$) 
in the sudden limit \cite{Mullen}. 
Similarly for a spin interacting with other spins in a lattice, 
its resonance window is defined by 
$|\gamma+\xi_i|\le\Delta_{win}/2$ for spin $i$, where
$\xi_i$ is the Zeeman energy of spin $i$ caused by 
the dipolar field from other spins. Since the dipolar field felt 
by spins is a distribution, only a small fraction of spins 
are in the resonance window at  any given moment.

With this physical picture in mind, we can 
write down the  evolution equation for the fraction $F$ of up-spins. 
If spins in the resonant window flip with probability $P_{win}$, 
we have
\begin{equation}\label{eq:evol}
\frac{dF}{d\gamma}=(1-2F)D(-\gamma,F)P_{win}\,,
\end{equation}
where  $D(\xi,F)$ is the normalized distribution of dipolar field
with the fraction $F$ of up-spins randomly located throughout 
the lattice, and $\gamma=\alpha t$ represents the sweeping field. The 
combination $(1-2F)D(-\gamma,F)$ is the difference between 
the fractions of up-spins and down-spins in the resonance window.
We want to solve Eq.(\ref{eq:evol}) with the initial condition
$F=0$, that is, all the spins point downwards at the beginning. 
The result  $F_{fin}=F(t\to\infty)$ is the fraction of up-spins
at the end of the sweep, or the normalized height $\delta M/2M_s$
of the quantum step between the two spin states. 
However, we need to first find what $P_{win}$ is, and how
to calculate the distribution function $D(\xi,F)$. 

\begin{figure}[!htb]
\begin{center}
\resizebox *{8.5cm}{6.57cm}{\includegraphics*{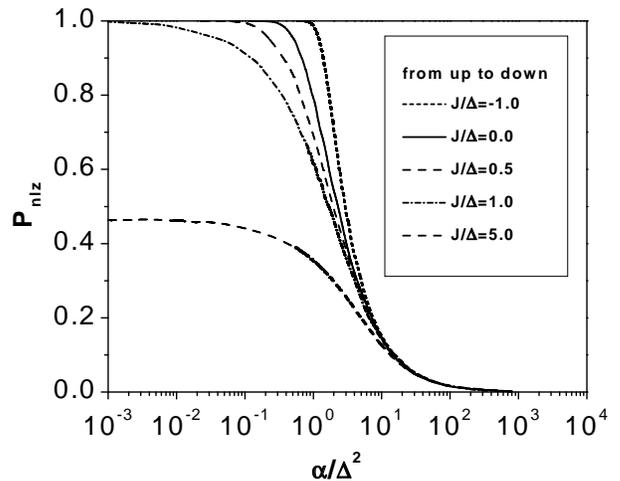}}
\end{center}
\caption{The flipping probability obtained with the 
nonlinear LZ model. $J/\Delta=0.0$ corresponds to the linear
LZ model. The flipping probability is suppressed for positive
$J$, and enhanced for negative $J$.}
\label{fig:nlz}
\end{figure}
Without the dipolar interaction, the flipping probability $P_{win}$ 
would be given by the LZ model, that is,
$P_{win}=P_{lz}=1-\exp(-\frac{\pi\Delta^2}{2\alpha})$.
With the dipolar interaction, it is no longer trivial to calculate
$P_{win}$.
During the tunneling time defined by $\Delta_{win}/\alpha$,
a spin inside the resonance window feels two kinds of dipolar fields:
one from spins outside the window, the other from spins inside the window
and trying to flip together. The former remains static during the short-time 
flipping process; it merely 
defines the position of the resonance window and does not
affect the flipping probability. In contrast, the latter is
changing with time, and will strongly affect $P_{win}$.

To account for this effect, we use a mean-field theory, treating each 
spin inside the resonance window equally. The interaction
is described by adding into the LZ model a nonlinear term, $J\bar{s}$,
where $J$ is the average coupling constant between spins 
and $\bar{s}$ is the average spin. 
This nonlinear LZ model was first proposed in the context
of Bose-Einstein condensation\cite{wuniu}. The coupling constant $J$
is proportional to the average  fraction of spins 
in the resonant window and is calculated as 
$J/\Delta=J_0\sqrt{1+2\alpha/\Delta^2} D(-\gamma,F)(1-2F)^2$,
where $J_0=\sum_jd(\vec{r}_j)$ is the dipolar field when
all the spins point in the same direction.

The probability $P_{win}$ is then given by the flipping probability 
$P_{nlz}$ obtained with this nonlinear model, which has been 
solved numerically and plotted in Fig.\ref{fig:nlz}.
The flipping probability is suppressed for positive 
$J/\Delta$ and increased for negative $J/\Delta$,
compared with the linear LZ probability. Furthermore,
we have found that the nonlinear LZ flipping probability
depends on only two parameters, 
$P_{nlz}=P_{nlz}(\alpha/\Delta^2,J/\Delta)$, and it has
an approximate expression \cite{wuniu,long}, 
\begin{equation}
P_{nlz}^{-1}= P_{lz}^{-1}+\frac{\sqrt{2}}{\pi}
\frac{J}{\Delta}\sqrt{P_{nlz}}.
\end{equation}

The remaining  task is to calculate the distribution of 
local fields. The dipolar field felt by a spin in the lattice, 
consists of two parts: one is the demagnetization
field from the spins very far away; the other from the neighboring
spins inside a ball $B_r$ of radius $r\sim (E_D/\Delta)^{1/3}$.
Since the demagnetization $\xi_{dm}$ is contributed by the distant spins,
it is independent of the lattice structure and only depends on
the sample shape and the fraction of up-spins. Our calculation
shows that $\xi_{dm}= 2CE_D(2F-1)$, where the constant $C$ is called
the shape coefficient and can be calculated theoretically\cite{ccc}.
On the other hand, the dipolar field from the neighboring spins
is a distribution depending on the lattice structure.
With its center shifted by the demagnetization field, 
the overall distribution function is
\be\label{eq:dist}
D(\xi,F)=\int\frac{{\rm d}k}{2\pi}\bar{D}(k,F)e^{ik(\xi-\xi_{dm})},
\ee
where 
$\bar{D}(k,F)=\Pi_{\vec{r}_j\neq 0}[(1-F)e^{-ik d(\vec{r}_j)}
+Fe^{ik d(\vec{r}_j)}]$\cite{long}.
The distribution functions calculated with Eq.(\ref{eq:dist})
are compared to a Monte-Carlo simulation in Fig. \ref{fig:dist};
there is an excellent agreement.    
\begin{figure}[!htb]
\begin{center}
\resizebox *{8.0cm}{6.18cm}{\includegraphics*{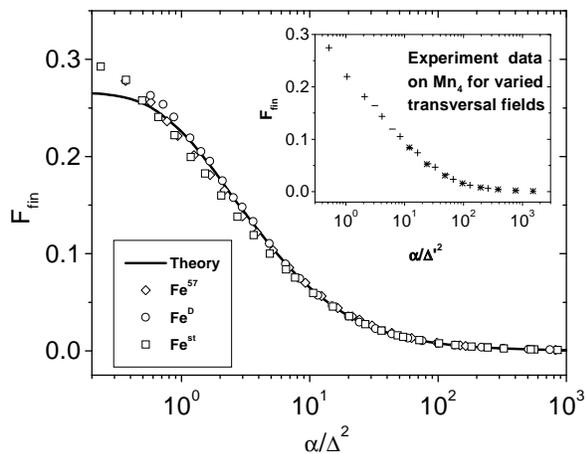}}
\end{center}
\caption{Comparison between our theory and  experiments.
The experimental data on Fe$_8$ isotopes from Ref.\cite{w1}, which have
different tunnel splittings $\Delta$, collapse
on the same curve, demonstrating the $\alpha/\Delta^2$
scaling law. The inset shows the collapse of the
data for Mn$_4$ Ref.\cite{newexp}, whose $\Delta$ is varied by changing
the transverse field. The slight deviation in the 
adiabatic regime $\alpha/\Delta^2<0.5$ is likely caused
by the ``hole digging'' mechanism \cite{see}.}
\label{fig:scaling}
\end{figure}

The above discussions indicate that the parameter $J_0$          
also consists of two parts, $J_0=\sum_{B_r}d(\vec{r})+2CE_D$.
The first part only relates to the crystal structure; for the 
triclinic Fe$_8$, centered tetragonal Mn$_{12}$, and hexagonal Mn$_4$
its value is $3.98E_D$,$1.15E_D$ and $12.63E_D$, respectively. 

Combining Eqs.(1-3), we can integrate the evolution 
equation (\ref{eq:evol}) with the initial condition $F=0$.
The results for the tunneling between the two spin states
$S=\pm 10$ of Fe$_8$ are plotted in Fig. \ref{fig:scaling}. 
With the horizontal axis taken as $\alpha/\Delta^2$, 
the experimental data for three different isotopes of Fe$_8$ 
collapse  onto the curve given by our theory.
This ``collapse'' is expected: $\Delta$ and $\alpha$ enter 
Eq.(\ref{eq:evol}) only in the combination of $\alpha/\Delta^2$
through $P_{nlz}$. This remarkable scaling law,
along with the excellent agreement of our theory with the experiments,
strongly supports our previous argument that nuclear 
spins do not appreciably  affect the flipping dynamics of the 
molecular spins, except for renormalizing the tunnel splitting 
through hyperfine coupling. This scaling law is
further confirmed by a set of new experimental data
on a different system, Mn$_4$ with $S=\frac 9 2$ \cite{newexp}.
By changing the transversal magnetic field
from $0$T to $0.085$T, the tunnel splitting of Mn$_4$
is varied almost by an order of magnitude.
Nevertheless, these data collapse perfectly onto a single curve
(inset of Fig. \ref{fig:scaling}).

One more important feature in 
Figs. \ref{fig:split}\&\ref{fig:scaling} is
the strong suppression of the
quantum step height, compared with the predictions of the LZ model.
Two physical mechanisms are behind the suppression. 
One is the ferromagnetic blocking between spins in the resonance
window. Due to the sample shape\cite{cccfe} and with
the shortest lattice vector being along the easy axis,
the dipolar interaction between spins in Fe$_8$ is very much 
ferromagnetic, yielding a positive coupling constant $J$. 
As seen in Fig. \ref{fig:nlz}, the flipping probability  
$P_{nlz}$ is suppressed from  $P_{lz}$ for this case.
This effect is relatively more significant in the fast sweeping regime,
where there are more spins in the resonance window due to
the broadened window width and narrow dipolar field distribution.
Our calculation finds suppression of up to 13\% due to this mechanism.  
However, this does not account for
all the suppression, especially for the slow sweeping limit
where the resonant window is narrower and the distribution function
is wider.

The other mechanism is the shuffling of
spins across the spectrum of the dipolar distribution
$D(\xi,F)$. As other spins flip, the dipolar field
$\xi_i$ felt by spin $i$ is altered and thus gets shuffled
to a different part of the spectrum. In particular, many spins which 
are yet to be brought into resonance can get shuffled into the 
swept part of the spectrum, losing their chances of flipping.
This is confirmed by our Monte-Carlo simulation\cite{long},
where the position of the resonance window is updated after
the spins in the window are flipped with probability $P_{win}$.
In Fig. \ref{fig:dist}, we show how the dipolar distribution
function changes with the sweeping field in
one simulation with $P_{win}=1$. Many spins in the main
peak are shuffled into the two right peaks. This dominant 
shuffling to the right is related to the largely ferromagnetic
character of the dipolar interaction between spins.
A careful tracking in our simulation shows that 
about 50\% of the spins are never brought into resonance 
and flip zero times, 28\% flip once, and 12\% flip twice. 
This shuffling mechanism gives an intuitive picture of the 
physics hidden in the evolution equation (\ref{eq:evol}).
\begin{figure}[!htb]
\begin{center}
\resizebox *{8.0cm}{6.18cm}{\includegraphics*{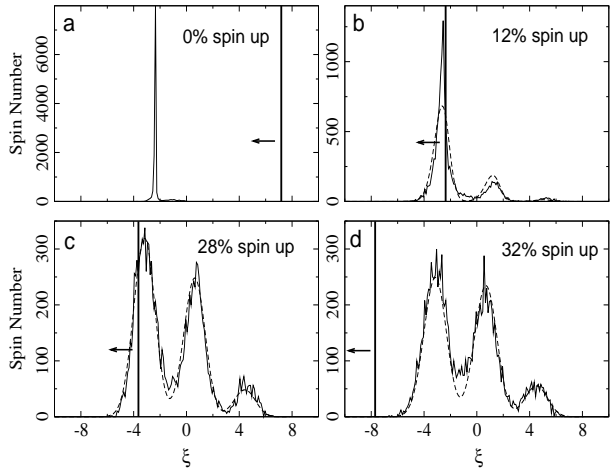}}
\end{center}
\caption{Change of the internal field distribution
in a sweeping field. $E_D\Omega_0=1, \Delta_{win}=0.1, P_{win}=1.$
The broadened vertical line represents the resonance
window. The dashed lines are calculated from Eq.(\ref{eq:dist}),
showing an  excellent agreement with the solid line.}
\label{fig:dist}
\end{figure}

\begin{figure}[!htb]
\begin{center}
\resizebox *{8.0cm}{6.18cm}{\includegraphics*{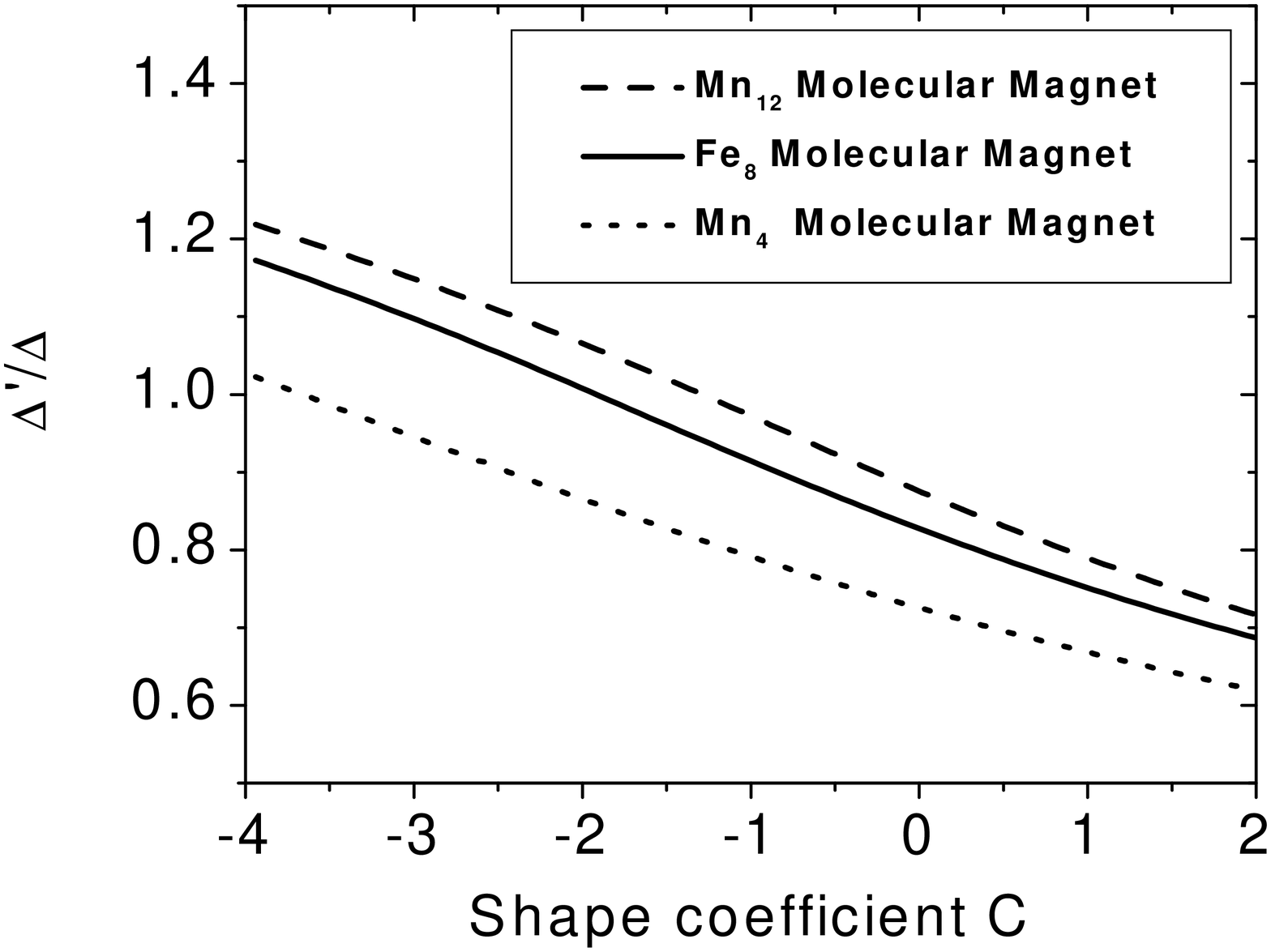}}
\end{center}
\caption{The dependence of 
$\Delta'/\Delta$ on the shape coefficient. $\Delta$ is
the true tunnel splitting; $\Delta'$ is the tunnel splitting
measured by the LZ method\cite{lzsci}.}
\label{fig:dcscal}
\end{figure}

With our theory, we can accurately determine the tunnel splitting 
$\Delta$ of a single molecule from the  measured  quantum 
step height. The LZ method \cite{lzsci,newexp} has been
used to accomplish this, extracting an effective tunnel 
splitting $\Delta'$ from a step height with 
$F_{fin}=1-\exp(-\pi\Delta'^2/2\alpha)$. However, the effective 
splitting $\Delta'$ is not necessarily the true splitting
$\Delta$ as the LZ model is inadequate to give the
correct step height.

Let us consider the fast sweeping limit,
where the magnetization is very small. In this case, 
the dipolar field distribution Eq.(3) is  a Lorentzian
centered at  $(\sum_{B_r}d(\vec{r})+2CE_D)(2F-1)$  with width
$\frac{8\pi^2}{3^{5/2}} E_D (2F) $\cite{loz}. Then,
using the new  variables $f=F/P_{lz}, x=\gamma/E_DP_{lz}$, 
one can rewrite Eq.(\ref{eq:evol}) 
and immediately notice that the evolution equation in terms of
$f$ and $x$ is independent of $P_{lz}$ and $E_D$.
It means that, in the regime $\alpha \to \infty$, the 
ratio $F_{fin}/P_{lz}$ $ = $ $(\Delta'/\Delta)^2$ tends to  a constant
depending only on $C$ and  $J_0/E_D$, which describe the 
geometry of the sample: its shape and lattice structure. 
This asymptotic relation explains why the $\Delta'$ is not 
necessarily the true $\Delta$ of a single molecule.
We have calculated the ratio $\Delta'/\Delta$ and its dependence
on the shape coefficient $C$ for  
three different molecular magnets Fe$_8$, Mn$_{12}$\cite{mn12},
and Mn$_4$ as shown in Fig. \ref{fig:dcscal}.
With this, one can obtain the real tunnel splitting from the 
corresponding step height given the shape of the sample.
For the Fe$_8$ sample used in the experiments
\cite{w1} the shape coefficient $C=1.4$\cite{cccfe}, which yields
$\Delta'/\Delta \simeq 0.73$.

We thank Wolfgang Wernsdorfer for stimulating discussion.
This project is  supported by the NSF of US, the Welch Foundation in
Texas, and the NNSF of China.

\vskip5pt
\hrule


\vskip5pt
\begin{thebibliography}{99}
\bibitem{mn12}J.R. Friedman {\it et al.}, 
Phys. Rev. Lett., {\bf 76}, 3830 (1996); 
C. Sangregorio {\it et al.}, ibid., 
{\bf 78}, 4645 (1997); L. Thomas {\it et al.}, Nature, {\bf 383}, 
145 (1996); J.M. Hernandez {\it et al.},
Europhys. Lett. {\bf 35} 301 (1996).

\bibitem{ljk} W. Wernsdorfer, Adv. Chem. Phys., {\bf 118}, 99 (2001)
(references therein).

\bibitem{all}L. Gunther and B. Barbara, 
{\it Quantum Tunneling of Magnetization}(Kluwer Academic, London, 1995);
A.L. Barra {\it et al.}, Europhys. Lett. {\bf 35}, 133 (1996);
A. Garg, ibid. {\bf 22}, 205 (1993);
A. Garg, Phys. Rev. B {\bf 51}, 15161 (1995);
F. Luis {\it et al.}, ibid., {\bf 57}, 505 (1998); 
D.A. Garanin, and E. M. Chudnovsky, ibid. {\bf 59}, 3671 (1999);
D.A. Garanin {\it et al.}, ibid. {\bf 61}, 12204 (2000);
M.N. Leuenberger and Daniel Loss, ibid. {\bf 61} 1286(2000);
E.M. Chudnovsky, {\it Macroscopic Quantum Tunneling of
the Magnetic Moment} (Cambridge University Press, 1998).

\bibitem{lzsci}W. Wernsdorfer and R. Sessoli, Science, 
{\bf 284}, 133 (1999).

\bibitem{qc}M.N. Leuenberger and D. Loss, Nature, {\bf 410}, 789 (2001).

\bibitem{w1}W. Wernsdorfer {\it et al.}, 
Europhys. Lett., {\bf 50}, 552 (2000).

\bibitem{lz}  L.D. Landau, Phys. Z. Sowjetunion {\bf 2}, 46 (1932); C.
Zener, Proc. R. Soc. London, Ser, A {\bf 137}, 696 (1932).

\bibitem{newexp}W. Wernsdorfer {\it et al.}, cond-mat/0109067.

\bibitem{exp}J.A.A.J. Perenboom {\it et al.}, 
Phys. Rev. B {\bf 58} 330(1998); 
Y. Zhong {\it et al.}, ibid. {\bf 62} R9256(2000);
Z.H. Jang {\it et al.}, Phys. Rev. Lett. {\bf 84} 2977(2000);
L. Bokacheva {\it et al.}, ibid. {\bf 85} 4803(2000).


\bibitem{stamp}N.V. Prokof'ev and P.C.E. Stamp, Phys. Rev. Lett. {\bf 80},
5794 (1998).

\bibitem{w2}W. Wernsdorfer {\it et al.},
J. Appl. Phys. {\bf 87} 5481 (2000).

\bibitem{ss}This is for the special case, the tunneling between
the two states $S_z=\pm S$. In general, for the tunneling 
between $S_z=S_1$ and $S_z=S_2$, we should replace $2S$ by
$|S_1-S_2|$.

\bibitem{Mullen} K. Mullen {\it et al.}, Phys. Rev. Lett. {\bf 62},
2543 (1989); Qian Niu and M.G. Raizen, Phys. Rev. Lett. {\bf 80},
3491 (1998). In this letter, we interpolate these two limits
by $\Delta_{win}=\sqrt{\Delta^2+2\alpha}$.




\bibitem{wuniu}B. Wu and Q. Niu, Phys. Rev. A {\bf 61}, 023402 (2000);
J. Liu {\it et al.}, {\it Quan-ph/0105140(2001)}.



\bibitem{long}J. Liu {\it et al.},
preprint, {\it Spin Tunneling of Molecular Magnets in
Sweeping Magnetic Fields}, in preparation, (2001).

\bibitem{ccc} For an ellipsoid with three axises $a$,$b$ and $c$, the 
shape coefficient $C=2\pi(1/3-R_g)$. Here the demagnetization
factor $R_g = 0.5abc \int_0^\infty \frac{dx}{(x+a^2)
\sqrt{(x+a^2)(x+b^2)(x+c^2)}}$. 


\bibitem{see}
For the experiment on the "hole-digging" see, 
W. Wernsdorfer {\it et al.},
Phys. Rev. Lett. {\bf 82} 3903 (1999); 
W. Wernsdorfer {\it et al.}, ibid. {\bf 84} 2965 (2000).
For the Monte-Carlo simulations, refer to,
T. Ohm {\it et al.}, Euro. Phys. J. B {\bf 6}, 195 (1998);
A. Cuccoli {\it et al.}, Euro. Phys. J. B {\bf 12}, 39 (1999);
J.J. Alonso and Julio F. Fernandez,
Phys. Rev. Lett. {\bf 87} 097205 (2001).



\bibitem{loz}A. Abragam, {\it Principle of Nuclear Magnetism}
(Oxford University Press 1961). It can also be derived from Eq.(\ref{eq:dist}).

\bibitem{cccfe}
From private  communication with Wernsdorfer,
in  the experiment of Fe$_8$ isotopes, the sample shape is $a=1$
(easy axis direction), $b=0.7$ and $c=0.2$. Using the formula in
\cite{ccc}, we have $C=1.4$.

\bibitem{mn}Here we only consider a perfect crystal, ignoring
effects induced by the dislocation (see cond-mat/0105518).

\end{thebibliography}
\end{document}